\documentclass[preprint,aps,prd,showpacs,floatfix]{revtex4-1}
\usepackage{times}
\usepackage{amsthm}
\usepackage{graphicx}
\usepackage{amsmath}
\usepackage{amssymb}
\usepackage{amsfonts}
\usepackage[english]{babel}
\usepackage{epstopdf}
\usepackage{multirow,makecell,array}
\usepackage{mathtools}
\usepackage{epstopdf}
\usepackage{color}
\usepackage{longtable}
\newcommand{\be}{\begin{eqnarray}}
\newcommand{\ee}{\end{eqnarray}}

\newcommand{\phz}{\phantom{0}}

\newcommand{\ket}[1]{|#1\rangle}

\newcommand{\matrixel}[3]{\langle #1 | #2 | #3 \rangle}
\newcommand{\calH}{{\cal H}}
\newcommand{\bcalH}{{\mbox{\boldmath$\cal H$}}}

\newcommand{\bfr}{{\bf r}}

\newcommand{\bfp}{{\bf p}}

\newcommand{\balpha}{{\mbox{\boldmath$\alpha$}}}
\newcommand{\bnabla}{{\mbox{\boldmath$\nabla$}}}
\newcommand{\dgrec}{\Delta g_{\mathrm{rec}}}
\newcommand{\dgrecBreit}{\Delta g_{\mathrm{Breit}}}
\newcommand{\dgrecQED}{\Delta g_{\mathrm{QED}}}
\newcommand{\FBreit}{F_{\mathrm{Breit}}}
\newcommand{\FQED}{F_{\mathrm{QED}}}
\newcommand{\Frec}{F_{\mathrm{rec}}}
\newcommand{\dgrece}{\Delta g_{\mathrm{nonmagn}}}
\newcommand{\dgrecm}{\Delta g_{\mathrm{magn}}}
\newcommand{\dgrecelo}{\dgrece^{(0)}}
\newcommand{\dgrecmlo}{\dgrecm^{(0)}}
\newcommand{\dgrecez}{\dgrece^{(1)}}
\newcommand{\dgrecmz}{\dgrecm^{(1)}}

\newcommand{\dgrecmzz}{\dgrecm^{(2)}}

\newcommand{\dgrecek}[1]{\dgrece^{(#1)}}
\newcommand{\dgrecmk}[1]{\dgrecm^{(#1)}}
\newcommand{\Hmagn}{H_{\mathrm{magn}}}
\newcommand{\HM}{H_{M}}
\newcommand{\HMmagn}{H_{M}^{\mathrm{magn}}}
\newcommand{\hD}{h_{\mathrm{D}}}
\newcommand{\Hint}{H_{\mathrm{int}}}
\newcommand{\HDCB}{H_{\mathrm{DCB}}}
\newcommand{\Vnucl}{V_{\mathrm{nucl}}}
\newcommand{\Vscr}{V_{\mathrm{scr}}}
\newcommand{\aZ}{\alpha Z}

\newcommand{\lstgr}{[(1s)^2\,(2s)^2\,2p]\,{}^2P_{1/2}}

\begin{document}
\thispagestyle{empty}

\title{Interelectronic-interaction contribution to the nuclear recoil effect\\ on the $g$ factor of boronlike ions}

\author{D.~A.~Glazov}
\author{A.~V.~Malyshev}
\author{V.~M.~Shabaev}
\author{I.~I.~Tupitsyn}
\affiliation{Department of Physics, St.~Petersburg State University, 
Universitetskaya 7/9, 199034 St.~Petersburg, Russia
}

%%%%%%%%%%%%%%%%%%%%%%%%%%%%%%%%%%%%%%%%%%%%%%%%%%%%%%%%%%%%%%%%%%%%%%%%
% 
\begin{abstract}
The nuclear recoil effect on the ground-state $g$ factor of highly charged boronlike ions is considered within the relativistic formalism. The interelectronic-interaction contribution is evaluated in the Breit approximation employing two independent approaches: the second-order perturbation theory and the configuration-interaction Dirac-Fock-Sturm method. The uncertainty of the nuclear recoil $g$-factor contributions is significantly reduced, especially for low- and middle-$Z$ ions.
\end{abstract}
% 
%%%%%%%%%%%%%%%%%%%%%%%%%%%%%%%%%%%%%%%%%%%%%%%%%%%%%%%%%%%%%%%%%%%%%%%%

\maketitle
%
%%%%%%%%%%%%%%%%%%%%%%%%%%%%%%%%%%%%%%%%%%%%%%%%%%%%%%%%%%%%%%%%%%%%%%%%
% 
\section{Introduction}\label{sec:intro}
% 
%%%%%%%%%%%%%%%%%%%%%%%%%%%%%%%%%%%%%%%%%%%%%%%%%%%%%%%%%%%%%%%%%%%%%%%%
%
The bound-electron $g$ factor proved to be a versatile tool for fundamental physics \cite{shabaev:15:jpcrd,shabaev:18:hi}. In particular, the electron mass has been determined with the best precision to date from the $g$-factor studies in H-like carbon and silicon \cite{sturm:14:n,CODATA14}. Few-electron ions, in particular, Li-like and B-like, were found to be indispensable for deeper insights into the bound-state quantum electrodynamics (QED) effects. Independent determination of the fine structure constant $\alpha$ has been proposed on the basis of the specific difference of the $g$-factor values of H-like, Li-like, and B-like ions of the same isotope \cite{shabaev:06:prl,yerokhin:16:prl}. Following the high-precision $g$-factor measurements in H-like ions \cite{haeffner:00:prl,verdu:04:prl,sturm:11:prl,sturm:13:pra}, the experiments with Li-like silicon \cite{wagner:13:prl,glazov:19:prl} and calcium \cite{koehler:16:nc} have provided the most stringent tests of the many-electron bound-state QED effects in the presence of external magnetic field \cite{volotka:14:prl,yerokhin:17:pra,glazov:19:prl}. Finally, the recent measurements in B-like argon \cite{arapoglou:19:prl,egl:19:prl} within the ALPHATRAP project \cite{sturm:19:epjst} extend these tests to $p$ states. While the experimental value of the $g$ factor of the ground $2p_{1/2}$ state is known to the ppb precision \cite{arapoglou:19:prl}, for the first excited $2p_{3/2}$ state the uncertainty is about $10^{-4}$ \cite{egl:19:prl}. 
% In this case, the $g$ factor has been determined by resolving the Zeeman components of the optical fine-structure transition, similar to the previous measurement \cite{soriaorts:07:pra} but an order of magnitude more precise. 
Nevertheless, in the anticipated experiments the $2p_{3/2}$-state $g$ factor can be measured with the ppb precision as well \cite{lindenfels:13:pra,vogel:18:ap,sturm:19:epjst}. The experimental values for both states are in perfect agreement with theory \cite{soriaorts:07:pra,glazov:13:ps,agababaev:18:jpcs,arapoglou:19:prl,agababaev:19:xrs}.

Studies of the relativistic nuclear recoil effect can provide an access to the bound-state QED effects beyond the Furry picture (i.e., beyond the external field approximation) in the strong coupling regime. The $g$ factor of heavy ions is one of the best targets of these studies in view of anticipated experimental and theoretical accuracy \cite{malyshev:17:jetpl}. The recent measurement of the $g$-factor isotope shift in Li-like calcium \cite{koehler:16:nc} demonstrated the feasibility of the corresponding experimental investigations. In this regard, we performed systematic calculations of the nuclear recoil correction to the $g$ factor of Li-like ions in the range $Z=3$--$92$ \cite{shabaev:17:prl,shabaev:18:pra}. These calculations comprise the non-trivial QED and interelectronic-interaction contributions of the first order in the electron-to-nucleus mass ratio $m/M$. In the course of these investigations, we revealed the incompleteness of the previous calculations~\cite{hegstrom:73:pra,hegstrom:75:pra,yan:01:prl,yan:02:jpb}.

We also extended these calculations to the ground state of B-like ions: the leading relativistic effects were evaluated for $Z=10$--$20$ in Ref.~\cite{glazov:18:os}, while in Ref.~\cite{aleksandrov:18:pra} the contributions of zeroth order in $1/Z$ (the independent-electron approximation) were evaluated to all orders in $\aZ$ for $Z=20$--$92$. In both works, the interelectronic interaction was considered within the first-order perturbation theory. In order to take into account the higher-order contributions to some extent, various effective screening potentials were introduced in the zeroth-order approximation. However, it was observed that the zeroth-order screening effect strongly underestimates the evaluated first-order correction. Based on simple considerations, it was concluded that the second-order correction cannot be seized by the first-order calculation with any screening potential. For this reason, the uncertainty due to uncalculated second-order correction was estimated by the first-to-zeroth order ratio multiplied by the first-order correction obtained in the pure Coulomb potential. This uncertainty dominates for low-$Z$ and middle-$Z$ ions, even exceeding the total non-trivial QED part for $Z<50$. In particular, for B-like argon this uncertainty is only three times smaller than the total theoretical $g$-factor uncertainty determined by the two-electron QED and higher-order interelectronic-interaction contributions \cite{arapoglou:19:prl}. These contributions can be significantly improved within the methods previously developed for Li-like ions \cite{volotka:14:prl,yerokhin:17:pra,glazov:19:prl}. In this case, the total theoretical uncertainty would be determined by the nuclear recoil effect, i.e., by the its second- and higher-order interelectronic-interaction corrections. 

In the present work, we improve significantly the theoretical accuracy of the nuclear recoil effect on the $g$ factor of B-like ions. The total recoil contribution can be divided into the so-called magnetic and nonmagnetic parts, the first one being much larger than the second one, except for the very heavy ions. In order to eliminate the dominant uncertainty, the first-order perturbation-theory results are complemented by the second-order correction to the magnetic part. This contribution is found to be significant. As we expected, it is beyond the spread of the first-order results obtained with different screening potentials. Moreover, the magnetic part is calculated independently within the large-scale configuration-interaction Dirac-Fock-Sturm (CI-DFS) method. The values obtained within the two methods are found in good agreement. As a result, we present the most accurate to date values of the nuclear recoil effect on the $g$ factor of B-like ions in the range $Z=10$--$92$. In particular, for the B-like argon ion the uncertainty of the nuclear recoil correction is reduced by a factor of 10.

Relativistic units ($\hbar = 1$, $c = 1$) and Heaviside charge unit [$\alpha=e^2/(4\pi)$, $e<0$] are employed throughout the paper, $\mu_0=|e|/(2m)$ denotes the Bohr magneton.
%
%%%%%%%%%%%%%%%%%%%%%%%%%%%%%%%%%%%%%%%%%%%%%%%%%%%%%%%%%%%%%%%%%%%%%%%%
% 
\section{Theoretical methods}
\label{sec:formulas}
%
%%%%%%%%%%%%%%%%%%%%%%%%%%%%%%%%%%%%%%%%%%%%%%%%%%%%%%%%%%%%%%%%%%%%%%%%
% 
Fully relativistic (QED) theory of the nuclear recoil effect in the first order in the electron-to-nucleus mass ratio $m/M$ was developed in Ref.~\cite{shabaev:01:pra}. The formulas obtained in this work were used to evaluate to all orders in $\alpha Z$ the nuclear recoil correction to the $g$ factor of the $1s$, $2s$, and $2p_j$ states~\cite{shabaev:02:prl,shabaev:17:prl,malyshev:17:jetpl,shabaev:18:pra,aleksandrov:18:pra,malyshev:19:}. At the same time, the leading relativistic nuclear recoil contributions can be evaluated with the effective four-component operators~\cite{shabaev:17:prl} derived from the QED formalism of Ref.~\cite{shabaev:01:pra}. In the present work we consider only this part, which can be termed as the Breit approximation for the nuclear recoil effect. We note, that the nuclear recoil effect on the $g$ factor of atomic systems can be treated also within the framework of the two-component approach (see, e.g., Ref.~\cite{wienczek:14:pra} and references therein). This approach allows one to evaluate the lowest-order relativistic contributions, while the nonrelativistic operator was derived by Phillips~\cite{phillips:49:pr}.

The nuclear recoil correction to the atomic $g$ factor within the Breit approximation is represented by the two effective relativistic operators~\cite{shabaev:01:pra,shabaev:17:prl}, 
\begin{align}
\label{eq:HMmagn}
  \HMmagn &= - \mu_0 \calH \frac{m}{M}
    \sum_{i,j} \left\{ \left[ \bfr_i\times \bfp_j \right]_z 
    - \frac{\alpha Z}{2r_j} \left[ \bfr_i\times\left( \balpha_j
    + \frac{(\balpha_j\cdot\bfr_j)\bfr_j}{r_j^2}\right) \right]_z
  \right\}
\,,\\
\label{eq:HM}
  \HM &= \frac{1}{2M} \sum_{i,j} \left[\bfp_i\cdot \bfp_j
  - \frac{\alpha Z}{r_i} \left( \balpha_i + \frac{(\balpha_i\cdot\bfr_i)\bfr_i}{r_i^2} \right)
  \cdot\bfp_j\right]
\,,
\end{align} 
where the external magnetic field $\bcalH$ is assumed to be directed along the $z$ axis, $\balpha$ is the vector of the Dirac matrices, and the summations run over all electrons of the system.
The contribution of $\HMmagn$ is given by the average value,
\begin{align}
\label{eq:dgrecm}
  \dgrecm &= \frac{1}{\mu_0 \calH M_J} \matrixel{A}{\HMmagn}{A} 
\,,
\end{align} 
where $\ket{A}$ is the many-electron wave function of the state under consideration with the total angular momentum projection $M_J$ on the $z$ axis.
The contribution of $\HM$ is given by the following perturbation theory expression, 
% of the first orders in $\HM$ and $\Hmagn$,
%
\begin{align}
\label{eq:dgrece}
  \dgrece &= \frac{2}{\mu_0 \calH M_J} \, {\sum_{N}}' \, 
    \frac{\matrixel{A}{\HM}{N}\matrixel{N}{\Hmagn}{A}} {E_A - E_N}
\,,
\end{align} 
where the summation runs over the complete spectrum of the many-electron states $|N\rangle$ including the single-particle negative-energy excitations. The prime here and below indicates that the terms with $E_N=E_A$ are excluded from the summation. The external-magnetic-field interaction operator $\Hmagn$ is
\begin{equation}
  \Hmagn = \mu_0 \calH m \, \sum_{j} \left[ \bfr_j \times \balpha_j \right]_z
\,.
\end{equation}
Due to the relativistic origin of the nonmagnetic part, its contribution is generally much smaller than that of the magnetic part. This rule is violated for $s$ states, where the magnetic part also tends to zero in the nonrelativistic limit.

Within the independent-electron approximation, the many-electron wave function $\ket{A}$ of the ground $\lstgr$ state of boronlike ion is the Slater determinant constructed from the corresponding eigenfunctions of the Dirac Hamiltonian,
\begin{equation}
\label{eq:hD}
  \hD = -i \balpha \cdot \nabla + \beta m + V (r)
\,.
\end{equation}
The spherically symmetric binding potential $V(r)$ includes the nuclear potential $\Vnucl(r)$ and optionally an effective screening potential $\Vscr(r)$, which can be introduced to account approximately for the interelectronic interaction already in the zeroth order. 

In order to take into account the interelectronic interaction beyond the effective-potential approximation, we consider the Dirac-Coulomb-Breit (DCB) Hamiltonian,
\begin{equation}
\label{eq:HDCB}
  \HDCB = \Lambda_+ \left( \sum_j \hD(j) + \Hint \right) \Lambda_+
\,,
\end{equation}
where the Coulomb-Breit interaction operator $\Hint$ reads, 
\begin{equation}
\label{eq:Hint}
  \Hint = \alpha \sum_{i<j} \left[ \frac{1}{r_{ij}} - \frac{ {\balpha}_i \cdot {\balpha}_j }{r_{ij}}
    - \frac{1}{2} ( {\balpha}_i \cdot {\bnabla}_i ) ( {\balpha}_j \cdot {\bnabla}_j ) r_{ij} \right]
    - \sum_j \Vscr(r_j)
\,,
\end{equation}
$r_{ij}=|\bfr_i-\bfr_j|$, and the positive-energy-states projection operator $\Lambda_+$ is constructed as the product of the one-electron projectors defined with respect to $\hD$~(\ref{eq:hD}). The last term in Eq.~(\ref{eq:Hint}) is needed to restore the original DCB Hamiltonian in case the screening potential $\Vscr$ is introduced in $\hD$.

The formulas (\ref{eq:dgrecm}) and (\ref{eq:dgrece}) can be used to find the nuclear recoil contributions to the $g$ factor for some (approximate) solution $\ket{A}$ of the Dirac-Coulomb-Breit equation, $\HDCB \ket{A} = E_A \ket{A}$. For realization of this scheme, we employ the CI-DFS method~\cite{bratsev:77}, which has been successfully applied for a wide variety of electronic-structure calculations. In particular, the interelectronic-interaction contribution (without the recoil effect) to the $g$ factor of B-like ions was obtained within the CI-DFS method in Refs.~\cite{glazov:13:ps,shchepetnov:15:jpcs,arapoglou:19:prl}. The reference-state wave function $\ket{A}$ is found as a linear combination of the configuration-state functions in the basis of the Dirac-Fock and Dirac-Fock-Sturm orbitals. Then, the magnetic part $\dgrecm$ is simply calculated as the average value of $\HMmagn$~(\ref{eq:dgrecm}). Evaluation of the nonmagnetic part~(\ref{eq:dgrece}) within the CI-DFS method is more complicated and, at the same time, less important, due to its smallness. So, it is not considered in the present work.

Alternatively, the perturbation theory with respect to $\Hint$ can be used. In this way, the zeroth-order terms $\dgrecmlo$ and $\dgrecelo$ are obtained from Eqs.~(\ref{eq:dgrecm}) and (\ref{eq:dgrece}) with the wave functions $\ket{A}$ and $\ket{N}$ constructed as the Slater determinants within the independent-electron approximation. The first-order correction to the magnetic part reads,
\begin{equation} 
\label{eq:dgrecmz}
  \dgrecmz = \frac{2}{\mu_0 \calH M_J} \, {\sum_{N}^{(+)}}{\vphantom{\sum}}'
    \frac{\matrixel{A}{\HMmagn}{N}\matrixel{N}{\Hint}{A}}{E_A-E_N}
\,,
\end{equation}
where the plus sign over the sum indicates that the intermediate $\ket{N}$ states are constructed as the Slater determinants of the positive-energy one-electron states only. The first-order correction $\dgrecez$ to the nonmagnetic part is given by the somewhat lengthy formula, which has been given in a compact form in Ref.~\cite{aleksandrov:18:pra}. In that work, all contributions up to the first order in $\Hint$ --- $\dgrecmlo$, $\dgrecmz$, $\dgrecelo$, and $\dgrecez$ --- have been evaluated in the Coulomb and four different screening potentials. The unknown contributions of the second and higher orders largely determine the total uncertainty in the wide range of $Z$, up to $Z \approx 80$. In this work, we evaluate the second-order correction to the magnetic part of the nuclear recoil effect,
\begin{align}
\label{eq:dgrecmzz}
  \dgrecmzz = \frac{1}{\mu_0 \calH M_J} \,
    \left[ 2\,{\sum_{N_{1}N_{2}}^{(+)}}{\vphantom{\sum}}'
      \frac{\matrixel{A}{\HMmagn}{N_1}\matrixel{N_1}{\Hint}{N_2}\matrixel{N_2}{\Hint}{A}}{(E_A-E_{N_1})(E_A-E_{N_2})} \right. &
\nonumber\\
         - 2\,\matrixel{A}{\Hint}{A}\,{\sum_{N_{1}}^{(+)}}{\vphantom{\sum}}'
      \frac{\matrixel{A}{\HMmagn}{N_1}\matrixel{N_1}{\Hint}{A}}{(E_A-E_{N_1})^2} &
\nonumber\\
         + {\sum_{N_{1}N_{2}}^{(+)}}{\vphantom{\sum}}'
      \frac{\matrixel{A}{\Hint}{N_1}\matrixel{N_1}{\HMmagn}{N_2}\matrixel{N_2}{\Hint}{A}}{(E_A-E_{N_1})(E_A-E_{N_2})} &
\nonumber\\
\left.   - \matrixel{A}{\HMmagn}{A}\,{\sum_{N_{1}}^{(+)}}{\vphantom{\sum}}'
      \frac{\matrixel{A}{\Hint}{N_1}\matrixel{N_1}{\Hint}{A}}{(E_A-E_{N_1})^2}
    \right] &
\,.
\end{align}
This expression leads to a massive set of contributions (which can be also represented as diagrams) within the standard many-body perturbation theory (MBPT) approach, i.e., when the many-electron wave functions $\ket{A}$, $\ket{N_{1}}$ and $\ket{N_{2}}$ are expanded in terms of the one-electron functions (eigenfunctions of $h_D$). In order to overcome this problem we follow the approach developed in Ref.~\cite{glazov:17:nimb}. First, the finite basis set of the Slater determinants is constructed from the one-electron basis of the dual-kinetically-balanced B-splines \cite{sapirstein:96:jpb,shabaev:04:prl}. Then, $\dgrecmzz$ is evaluated straightforwardly according to Eq.~(\ref{eq:dgrecmzz}), while the many-electron matrix elements are reduced to the one- and two-electron ones by the computer code according to the well-known combinatorial algorithm.  

In Ref.~\cite{glazov:17:nimb} the use of the recursive formulation of the perturbation theory was demonstrated on top of the effective treatment of the many-electron matrix elements. Within this approach, the contributions of the second and higher orders in $\Hint$ were evaluated for lithiumlike ions \cite{shabaev:17:prl,shabaev:18:pra}. Extension of this approach to boronlike systems is in demand, however, it is hold back by the fast growth of the many-electron basis set with respect to the number of excitations taken into account. The presently considered second-order contribution $\dgrecmzz$ involves the single and double excitations only, which makes it accessible for calculation.
%
%%%%%%%%%%%%%%%%%%%%%%%%%%%%%%%%%%%%%%%%%%%%%%%%%%%%%%%%%%%%%%%%%%%%%%%%
% 
\section{Results and discussion}
\label{sec:results}
%
%%%%%%%%%%%%%%%%%%%%%%%%%%%%%%%%%%%%%%%%%%%%%%%%%%%%%%%%%%%%%%%%%%%%%%%%
% 
The Breit-approximation nuclear recoil contributions to the $g$ factor of boronlike ions have been evaluated to zeroth and first orders in the interelectronic interaction in Ref.~\cite{glazov:18:os} for $Z=10$--$20$ and in Ref.~\cite{aleksandrov:18:pra} for $Z=20$--$92$. These works used different sets of the screening potentials, however, the results were found to be in good agreement at the ``junction point'' $Z=20$~\cite{aleksandrov:18:pra}. In this work we recalculate these contributions in the range $Z=10$--$18$ with the same choice of the potentials as in Ref.~\cite{aleksandrov:18:pra}. We also extend to this range of $Z$ the calculations of the higher-order (in $\aZ$) two-electron contribution, which has been presented in Ref.~\cite{shchepetnov:15:jpcs} for $Z=18$ and in Ref.~\cite{aleksandrov:18:pra} for $Z=20$--$92$. The higher-order one-electron contribution has been evaluated recently in the range $Z=1$--$20$ for $1s$, $2s$, $2p_{1/2}$ and $2p_{3/2}$ states~\cite{malyshev:19:}. 

The main focus of the present work is on the magnetic nuclear recoil contribution beyond the first order of the perturbation theory. Table~\ref{tab:C_magn} displays the second-order correction $\dgrecmk{2}$ evaluated according to Eq.~(\ref{eq:dgrecmzz}) as described in the previous section. The calculations have been performed for the Coulomb and four different screening potentials --- core-Hartree (CH), Perdew-Zunger (PZ)~\cite{perdew:81:prb}, Kohn-Sham (KS)~\cite{kohn:65:pr}, and local Dirac-Fock (LDF)~\cite{shabaev:05:pra}.
We present the results in terms of the coefficient $C(\aZ)$,
\begin{equation}
\label{eq:C_def}
  \dgrecmk{2} = \frac{m}{M} \frac{C(\aZ)}{Z^2}  
\,,
\end{equation}
defined in line with the $A(\aZ)$ and $B(\aZ)$ coefficients in Refs.~\cite{glazov:18:os,aleksandrov:18:pra}. So, the total magnetic part can be found as,
\begin{equation}
\label{eq:dgrecm_k}
  \dgrecm = \dgrecmk{0} + \dgrecmk{1} + \dgrecmk{2}
    = \frac{m}{M} \left[ A(\aZ) + \frac{B(\aZ)}{Z} + \frac{C(\aZ)}{Z^2} \right]
\,.
\end{equation}
In Table~\ref{tab:magn_total} we present the individual terms in the brackets in Eq.~(\ref{eq:dgrecm_k}), the sum of the first two terms $A(\aZ) + {B(\aZ)}/{Z}$, and the sum of all three terms. The results obtained with the different potentials are given in the corresponding columns. The zeroth- and first-order coefficients $A$ and $B$ were calculated already in Refs.~\cite{glazov:18:os,aleksandrov:18:pra}, however, only the sum of the magnetic and nonmagnetic parts was given, except for $Z=18$~\cite{glazov:18:os}. 

The magnetic nuclear-recoil contribution has been also evaluated within the all-order multireference CI-DFS method. The employed one-electron basis set is $30s\ 31p\ 32d\ 33f\ 34g$. All single and double excitations from the $(1s)^2 (2s)^2 2p$ and $(1s)^2 (2p)^3$ configurations are included. The contribution of the triple excitations is estimated with the smaller basis set, $20s\ 21p\ 22d$. The results are presented in Table~\ref{tab:magn_total} in the lines labeled as ``CI-DFS''. We note that these values are not related to any zeroth-order potential, Coulomb or screening.

Five columns in Table~\ref{tab:magn_total} represent the different zeroth-order potentials for the perturbation-theory results, the last column shows the ``spread'' of the screening-potential values. It is found as a maximal absolute difference between any two of them, while the Coulomb value is not included. The ``spread'' in the ``CI-DFS'' line is the absolute difference between the CI-DFS and the second-order LDF values. We point out the following important observations:\\
1. The spread is getting smaller after each step of the perturbation theory.\\
2. The contribution of the next order is significantly larger than the spread at the previous step. So, the precaution, based on the first-order results, that this spread cannot seize the unknown higher orders is confirmed now for the second order as well.\\
3. The second-order results are much closer to the CI-DFS values than the first-order results (except for the Coulomb potential for $Z \lesssim 50$). The differences between the second-order and the CI-DFS values are larger than the second-order spread (in agreement with the previous clause), but smaller than the second-order term $C/Z^2$ itself. So, the perturbation series can be termed as ``convergent'' at this level.\\
4. The results for the Coulomb potential demonstrate poor convergence of the perturbation series for $Z \lesssim 50$. This is not surprising and can be explained by the near degeneracy of the $(1s)^2 (2s)^2 2p$ and $(1s)^2 (2p)^3$ configurations in the zeroth-order approximation.\\
Based on the above remarks, we take the CI-DFS values as the final ones for the magnetic part. The uncertainty ascribed to these values equals the CI-DFS$-$LDF difference (``spread'' in the CI-DFS line). We accept this rather conservative estimation for the following reasons. The CI-DFS uncertainty due to the basis size is currently estimated to be much smaller --- from a few units for low $Z$ to less than 1 for high $Z$ in the last presented digit. However, the observed peculiarities of the perturbation series suggest that convergence of the all-order method might also be worse than it seems. The complete investigation of this question requires further extensive computations. There are also other sources of the total uncertainty that are discussed below.

The nonmagnetic part of the nuclear recoil effect on the $g$ factor of boronlike ions is presented in Table~\ref{tab:nonmagn_total}. The zeroth- and first-order terms, $A(\aZ)$ and ${B(\aZ)}/{Z}$, are presented along with their sum. The total $g$-factor contribution of the nonmagnetic part is,
\begin{equation}
\label{eq:g_e_k}
  \dgrece = \dgrecek{0} + \dgrecek{1}
    = \frac{m}{M} \left[ A(\aZ) + \frac{B(\aZ)}{Z} \right]
\,.
\end{equation}
The values obtained with the LDF potential are chosen as the final results. We estimate the uncertainty due to the second- and higher-order interelectronic-interaction corrections as $\dgrecez \cdot (\dgrecez/\dgrecelo)$. The higher-order results obtained in this work for the magnetic part demonstrate that this recipe proposed in Refs.~\cite{glazov:18:os,aleksandrov:18:pra} can be considered reliable.

The nuclear recoil contribution within the Breit approximation is given by the sum of the magnetic and nonmagnetic parts,
\begin{equation}
\label{eq:m-nm}
  \dgrecBreit = \dgrecm + \dgrece
\,.
\end{equation}
The total nuclear recoil correction to the $g$ factor is given by the sum of the Breit and QED terms,
\begin{align} 
\label{eq:total}
  \dgrec &= \dgrecBreit + \dgrecQED
\,.
\end{align}
Here, $\dgrecQED$ comprises both the one- and two-electron higher-order (in $\aZ$) contributions. The Breit and QED terms are presented in Table~\ref{tab:total} along with their sum in terms of the function $F(\aZ)$ defined by,
\begin{equation} 
\label{eq:F_def}
  \Delta g = \frac{m}{M} F(\alpha Z)
\,.
\end{equation}
For $\FQED$ we take the values from Ref.~\cite{aleksandrov:18:pra} in the range $Z=20$--$92$. For lower $Z$ we evaluate it in this work within the same methods for the LDF screening potential. The finite-nuclear-size effect is partly taken into account for all contributions by using the corresponding potential $\Vnucl(r)$ in the Dirac Hamiltonian~(\ref{eq:hD}). The Fermi model is employed to describe the nuclear charge distribution, and the nuclear charge radii are taken from Ref.~\cite{angeli:13:adndt}. The uncertainty due to the finite-nuclear-size effect has been discussed in Ref.~\cite{aleksandrov:18:pra}, we follow the same algorithm here and add the uncertainties quadratically. For the QED part, the uncertainty is obtained by multiplying $\Delta g_\text{QED}$ by a factor of $2/Z$. 

In Table~\ref{tab:total_ions}, we present the nuclear recoil contribution to the $g$ factor of several B-like ions in the range $Z=10-60$. We left out the lead and uranium ions, since the results are not improved in comparison to the ones presented in Ref.~\cite{aleksandrov:18:pra}. The nuclear masses are taken from the Ame2012 compilation~\cite{wang:12:cpc} in accordance with Ref.~\cite{yerokhin:15:jpcrd}. The nuclear recoil correction to the $g$ factor of boronlike argon ${}^{40}_{18}$Ar$^{13+}$ is now $-9.174\,(19)\times 10^{-6}$, which is $10$ times more accurate than the previous value $-9.09\,(19)\times 10^{-6}$ \cite{glazov:18:os}. This result represents an important step towards improvement of the total theoretical $g$-factor value, which is in high demand in view of the recent high-precision measurement~\cite{arapoglou:19:prl}. The interelectronic-interaction and two-electron QED effects, which are presently the main sources of uncertainty, can be evaluated to a better accuracy within the approach developed previously for lithiumlike ions \cite{volotka:14:prl,yerokhin:17:pra,glazov:19:prl}.
%
%%%%%%%%%%%%%%%%%%%%%%%%%%%%%%%%%%%%%%%%%%%%%%%%%%%%%%%%%%%%%%%%%%%%%%%%
% 
\section{Conclusion}\label{sec:conclusion}
%
%%%%%%%%%%%%%%%%%%%%%%%%%%%%%%%%%%%%%%%%%%%%%%%%%%%%%%%%%%%%%%%%%%%%%%%%
% 
The interelectronic-interaction contribution to the nuclear recoil effect on the ground-state $g$ factor of boronlike ions has been considered within the Breit approximation. The second-order correction to the dominant magnetic part has been calculated within the perturbation theory starting from the Dirac equation with an effective screening potential. The all-order results are obtained within the configuration-interaction Dirac-Fock-Sturm method. The nonmagnetic part and the nontrivial QED contributions are taken into account as well. The results are presented for the wide range of the nuclear charge number $Z=10$--$92$, the uncertainty is significantly improved for low and middle $Z$ ions. These results are important in view of the planned and already performed high-precision $g$-factor measurements for boronlike ions.
%
%%%%%%%%%%%%%%%%%%%%%%%%%%%%%%%%%%%%%%%%%%%%%%%%%%%%%%%%%%%%%%%%%%%%%%%%
% 
\section*{Acknowledgments}
%
%%%%%%%%%%%%%%%%%%%%%%%%%%%%%%%%%%%%%%%%%%%%%%%%%%%%%%%%%%%%%%%%%%%%%%%%
% 
This work was supported by the Russian Science Foundation (Grant No. 17-12-01097).
%
%
%
%%%%%%%%%%%%%%%%%%%%%%%%%%%%%%%%%%%%%%%%%%%%%%%%%%%%%%%%%%%%%%%%%%%%%%%%
%
% \bibliography{database}

\begin{thebibliography}{99}
%
\bibitem{shabaev:15:jpcrd}
% Theory of Bound-Electron g Factor in Highly Charged Ions
V.~M.~Shabaev, D.~A.~Glazov, G.~Plunien, and A.~V.~Volotka,
J.~Phys.~Chem.~Ref.~Data~\textbf{44}, 031205~(2015).
%
\bibitem{shabaev:18:hi}
% Stringent tests of QED using highly charged ions
V.~M.~Shabaev, A.~I.~Bondarev, D.~A.~Glazov, M.~Y.~Kaygorodov, Y.~S.~Kozhedub, I.~A.~Maltsev, A.~V.~Malyshev, R.~V.~Popov, I.~I.~Tupitsyn, and N.~A.~Zubova,
Hyperfine Interact. \textbf{239}, 60 (2018).
% 
\bibitem{sturm:14:n}
% High-precision measurement of the atomic mass of the electron
S.~Sturm, F.~K\"ohler, J.~Zatorski, A.~Wagner, Z.~Harman, G.~Werth, W.~Quint, C.~H.~Keitel, and K.~Blaum,
Nature \textbf{506}, 467 (2014).
% 
\bibitem{CODATA14}
% CODATA recommended values of the fundamental physical constants: 2014
P.~J.~Mohr, D.~B.~Newell, and B.~N.~Taylor,
Rev.~Mod.~Phys.~\textbf{88}, 035009~(2016).
%
\bibitem{shabaev:06:prl}
% g factor of heavy ions: a new access to the fine structure constant
V.~M.~Shabaev, D.~A.~Glazov, N.~S.~Oreshkina, A.~V.~Volotka, G.~Plunien, H.-J.~Kluge, and W.~Quint,
Phys. Rev. Lett. \textbf{96}, 253002 (2006).
%
\bibitem{yerokhin:16:prl}
% g Factor of Light Ions for an Improved Determination of the Fine-Structure Constant
V.~A.~Yerokhin, E.~Berseneva, Z.~Harman, I.~I.~Tupitsyn, and C.~H.~Keitel,
Phys.~Rev.~Lett.~\textbf{116}, 100801~(2016).
% 
\bibitem{haeffner:00:prl}
% High-Accuracy Measurement of the Magnetic Moment Anomaly of the Electron Bound in Hydrogenlike Carbon
H.~H\"affner, T.~Beier, N.~Hermanspahn, \mbox{H.-J.}~Kluge, W.~Quint, S.~Stahl, J.~Verd\'u, and G.~Werth,
Phys. Rev. Lett. \textbf{85}, 5308 (2000).
%
\bibitem{verdu:04:prl}
% Electronic g Factor of Hydrogenlike Oxygen O^{7+}
J.~Verd\'u, S.~Djeki\'c, S.~Stahl, T.~Valenzuela, M.~Vogel, G.~Werth, T.~Beier, \mbox{H.-J.}~Kluge, and W.~Quint,
Phys. Rev. Lett. \textbf{92}, 093002 (2004).
%
\bibitem{sturm:11:prl}
% g Factor of Hydrogenlike 28^Si^13+
S.~Sturm, A.~Wagner, B.~Schabinger, J.~Zatorski, Z.~Harman, W.~Quint, G.~Werth, C.~H.~Keitel, and K.~Blaum,
Phys. Rev. Lett. \textbf{107}, 023002 (2011).
%
\bibitem{sturm:13:pra}
% g-factor measurement of hydrogenlike 28Si13+ as a challenge to QED calculations
S.~Sturm, A.~Wagner, M.~Kretzschmar, W.~Quint, G.~Werth, and K.~Blaum,
Phys. Rev. A \textbf{87}, 030501(R)~(2013).
%
\bibitem{wagner:13:prl}
% g factor of lithiumlike silicon 28^Si^11+
A.~Wagner, S.~Sturm, F.~K\"ohler, D.~A.~Glazov, A.~V.~Volotka, G.~Plunien, W.~Quint, G.~Werth, V.~M.~Shabaev, and K.~Blaum,
Phys. Rev. Lett. \textbf{110}, 033003 (2013).
% 
\bibitem{glazov:19:prl}
% g Factor of Lithiumlike Silicon: New Challenge to Bound-State QED
D.~A.~Glazov, F.~K\"ohler-Langes, A.~V.~Volotka, K.~Blaum, F.~Hei\ss e, G.~Plunien, W.~Quint, S.~Rau, V.~M.~Shabaev, S.~Sturm, and G.~Werth,
Phys. Rev. Lett. \textbf{123}, 173001 (2019).
%
\bibitem{koehler:16:nc}
% Isotope dependence of the Zeeman effect in lithium-like calcium
F.~K\"ohler, K.~Blaum, M.~Block, S.~Chenmarev, S.~Eliseev, D.~A.~Glazov, M.~Goncharov, J.~Hou, A.~Kracke, D.~A.~Nesterenko, Yu.~N.~Novikov, W.~Quint, E.~Minaya~Ramirez, V.~M.~Shabaev, S.~Sturm, A.~V.~Volotka, and G.~Werth,
Nature Communications~\textbf{7}, 10246~(2016).
%
\bibitem{volotka:14:prl}
% Many-Electron QED Corrections to the g Factor of Lithiumlike Ions
A.~V.~Volotka, D.~A.~Glazov, V.~M.~Shabaev, I.~I.~Tupitsyn, and G.~Plunien,
Phys. Rev. Lett. \textbf{112}, 253004 (2014).
%
\bibitem{yerokhin:17:pra}
% Electron-correlation effects in the g factor of light Li-like ions
V.~A.~Yerokhin, K.~Pachucki, M.~Puchalski, Z.~Harman, and C.~H.~Keitel,
Phys. Rev. A \textbf{95}, 062511 (2017).
%
\bibitem{arapoglou:19:prl}
% The g-factor of Boronlike Argon 40_Ar^13+
I.~Arapoglou, A.~Egl, M.~H\"ocker, T.~Sailer, B.~Tu, A.~Weigel, R.~Wolf, H.~Cakir, V.~A.~Yerokhin, N.~S.~Oreshkina, V.~A.~Agababaev, A.~V.~Volotka, D.~V.~Zinenko, D.~A.~Glazov, Z.~Harman, C.~H.~Keitel, S.~Sturm, and K.~Blaum,
Phys. Rev. Lett. \textbf{122}, 253001 (2019).
%
\bibitem{egl:19:prl}
% Application of the Continuous Stern-Gerlach Effect for Laser Spectroscopy of the $^{40}{\mathrm{Ar}}^{13+}$ Fine Structure in a Penning Trap
A.~Egl, I.~Arapoglou, M.~H{\"o}cker, K.~K{\"o}nig, T.~Ratajczyk, T.~Sailer, B.~Tu, A.~Weigel, K.~Blaum, W.~N{\"o}rtersh{\"a}user, and S.~Sturm,
Phys. Rev. Lett. \textbf{123}, 123001 (2019).
% 
\bibitem{sturm:19:epjst}
% The ALPHATRAP experiment
S.~Sturm, I.~Arapoglou, A.~Egl, M.~H{\"o}cker, S.~Kraemer, T.~Sailer, B.~Tu, A.~Weigel, R.~Wolf, J.~C.~L{\'o}pez-Urrutia, K.~Blaum,
Eur. Phys. J. Special Topics \textbf{227}, 1425 (2019).
%
\bibitem{lindenfels:13:pra}
% Experimental access to higher-order Zeeman effects by precision spectroscopy of highly charged ions in a Penning trap
D.~von~Lindenfels, M.~Wiesel, D.~A.~Glazov, A.~V.~Volotka, M.~M.~Sokolov, V.~M.~Shabaev, G.~Plunien, W.~Quint, G.~Birkl, A.~Martin, and M.~Vogel,
Phys. Rev. A \textbf{87}, 023412 (2013).
%
\bibitem{vogel:18:ap}
% Electron Magnetic Moment in Highly Charged Ions: The ARTEMIS Experiment
M.~Vogel, M.~S.~Ebrahimi, Z.~Guo, A.~Khodaparast, G.~Birkl, and W.~Quint,
Ann. Phys. (Berlin) \textbf{531}, 1800211 (2019).
%
\bibitem{soriaorts:07:pra}
% Zeeman splitting and g factor of the 1s^2 2s^2 2p ^2P_3∕2 and ^2P_1∕2 levels in Ar^13+
R.~Soria Orts, J.~R.~Crespo L\'opez-Urrutia, H.~Bruhns, A.~J.~Gonz\'alez Mart\'inez, Z.~Harman, U.~D.~Jentschura, C.~H.~Keitel, A.~Lapierre, H.~Tawara, I.~I.~Tupitsyn, J.~Ullrich, and A.~V.~Volotka,
Phys. Rev. A \textbf{76}, 052501 (2007).
%
\bibitem{glazov:13:ps}
% g factor of boron-like ions: ground and excited states
D.~A.~Glazov, A.~V.~Volotka, A.~A.~Schepetnov, M.~M.~Sokolov, V.~M.~Shabaev, I.~I.~Tupitsyn, and G.~Plunien,
Phys. Scr. \textbf{T156}, 014014 (2013).
% 
\bibitem{agababaev:18:jpcs}
% Ground-state g factor of middle-Z boronlike ions
V.~A.~Agababaev, D.~A.~Glazov, A.~V.~Volotka, D.~V.~Zinenko, V.~M.~Shabaev, and G.~Plunien,
J.~Phys. Conf.~Ser. \textbf{1138}, 012003 (2018).
% 
\bibitem{agababaev:19:xrs}
% g factor of the [(1s)^2 (2s)^2 2p]^2P_3/2 state of middle-Z boronlike ions
V.~A.~Agababaev, D.~A.~Glazov, A.~V.~Volotka, D.~V.~Zinenko, V.~M.~Shabaev, and G.~Plunien,
X-Ray Spectrometry, in press (2019).
%
\bibitem{malyshev:17:jetpl}
% Nuclear recoil effect on $g$ factor of heavy ions: prospects for tests of quantum electrodynamics in a new region
A.~V.~Malyshev, V.~M.~Shabaev, D.~A.~Glazov, and I.~I.~Tupitsyn,
Pisma Zh. Eksp. Teor. Fiz.~\textbf{106}, 731~(2017) [JETP~Lett.~\textbf{106}, 765~(2017)].
% 
\bibitem{shabaev:17:prl}
% Recoil Effect on the g Factor of Li-Like Ions
V.~M.~Shabaev, D.~A.~Glazov, A.~V.~Malyshev, and I.~I.~Tupitsyn,
Phys.~Rev.~Lett.~\textbf{119}, 263001~(2017).
%
\bibitem{shabaev:18:pra}
% Nuclear recoil effect on the g factor of highly charged Li-like ions
V.~M.~Shabaev, D.~A.~Glazov, A.~V.~Malyshev, and I.~I.~Tupitsyn,
Phys.~Rev.~A~\textbf{98}, 032512~(2018).
%
\bibitem{hegstrom:73:pra}
% Nuclear-mass and anomalous-moment corrections to the Hamiltonian for an atom in a constant external magnetic field
R.~A.~Hegstrom,
Phys. Rev. A \textbf{7}, 451 (1973).
%
\bibitem{hegstrom:75:pra}
% Magnetic moment of atomic lithium
R.~A.~Hegstrom,
Phys. Rev. A \textbf{11}, 421 (1975).
%
\bibitem{yan:01:prl}
Z.-C.~Yan,
Phys. Rev. Lett. \textbf{86}, 5683 (2001).
%
\bibitem{yan:02:jpb}
Z.-C.~Yan,
J. Phys. B \textbf{35}, 1885 (2002).
%
\bibitem{glazov:18:os}
% Nuclear recoil effect on the g factor of middle-Z boronlike ions
D.~A.~Glazov, A.~V.~Malyshev, V.~M.~Shabaev, and I.~I.~Tupitsyn,
Opt. Spectrosc. \textbf{124}, 457 (2018).
% 
\bibitem{aleksandrov:18:pra}
% Relativistic nuclear-recoil effect on the g factor of highly charged boronlike ions
I.~A.~Aleksandrov, D.~A.~Glazov, A.~V.~Malyshev, V.~M.~Shabaev, and I.~I.~Tupitsyn,
Phys. Rev. A \textbf{98}, 062521 (2018).
% 
\bibitem{wienczek:14:pra}  
A.~Wienczek, M.~Puchalski, and K.~Pachucki,
Phys. Rev. A \textbf{90}, 022508 (2014). 
%
\bibitem{phillips:49:pr}
M.~Phillips,
Phys. Rev. \textbf{76}, 1803 (1949).
%
\bibitem{shabaev:01:pra}
% QED theory of the nuclear recoil effect on the atomic g factor
V.~M.~Shabaev,
Phys. Rev. A \textbf{64}, 052104 (2001).
%
\bibitem{shabaev:02:prl}
% Recoil Correction to the Bound-Electron g Factor in H-Like Atoms to All Orders in aZ
V.~M. Shabaev and V.~A. Yerokhin,
Phys. Rev. Lett. \textbf{88}, 091801 (2002).
%
\bibitem{malyshev:19:}
% Nuclear recoil effect on the g factor of highly charged Li-like ions
A.~V.~Malyshev, D.~A.~Glazov, and V.~M.~Shabaev, 
arXiv:1911.03978.
% 
\bibitem{bratsev:77}
% Application of the Hartree-Fock method to calculation of relativistic atomic wave functions
V.~F.~Bratzev and G.~B.~Deyneka and I.~I.~Tupitsyn,
Izv. Akad. Nauk SSSR, Ser. Fiz. \textbf{41}, 2655 (1977) [Bull. Acad. Sci. USSR, Phys. Ser. \textbf{41}, 173] (1977).
%
\bibitem{shchepetnov:15:jpcs}
% Nuclear recoil correction to the g factor of boron-like argon
A.~A.~Shchepetnov, D.~A.~Glazov, A.~V.~Volotka, V.~M.~Shabaev, I.~I.~Tupitsyn, and G.~Plunien,
J.~Phys. Conf. Ser.~\textbf{583}, 012001~(2015).
%
\bibitem{glazov:17:nimb}
% Higher-order perturbative relativistic calculations for few-electron atoms and ions
D.~A.~Glazov, A.~V.~Malyshev, A.~V.~Volotka, V.~M.~Shabaev, I.~I.~Tupitsyn, and G.~Plunien,
Nucl. Instrum. Methods Phys. Res. B \textbf{408}, 46~(2017).
%
\bibitem{sapirstein:96:jpb}
% The use of basis splines in theoretical atomic physics
J.~Sapirstein and W.~R.~Johnson,
J. Phys. B \textbf{29}, 5213 (1996).
%
\bibitem{shabaev:04:prl}
% Dual Kinetic Balance Approach to Basis-Set Expansions for the Dirac Equation
V.~M.~Shabaev, I.~I.~Tupitsyn, V.~A.~Yerokhin, G.~Plunien, and G.~Soff,
Phys. Rev. Lett. \textbf{93}, 130405 (2004).
% 
\bibitem{perdew:81:prb}  
J.~P.~Perdew and A.~Zunger, 
Phys. Rev. B \textbf{23}, 5048 (1981).
% 
\bibitem{kohn:65:pr}
W.~Kohn and L.~J.~Sham, 
Phys. Rev. \textbf{140}, A1133 (1965).
%
\bibitem{shabaev:05:pra}
% Radiative and correlation effects on the parity-nonconserving transition amplitude in heavy alkali-metal atoms
V.~M.~Shabaev, I.~I.~Tupitsyn, K.~Pachucki, G.~Plunien, and V.~A.~Yerokhin,
Phys. Rev. A \textbf{72}, 062105 (2005).
%
\bibitem{angeli:13:adndt}
% A consistent set of nuclear rms charge radii: properties of the radius surface R(N;Z)
I.~Angeli and K.~P.~Marinova,
At. Data Nucl. Data Tables \textbf{99}, 69 (2013).
%
\bibitem{wang:12:cpc}
% The Ame2012 atomic mass evaluation
M.~Wang, G.~Audi, A.~H.~Wapstra, F.~G.~Kondev, M.~MacCormick, X.~Xu, and B.~Pfeiffer,
Chin.~Phys.~C \textbf{36}, 1603 (2012).
%
\bibitem{yerokhin:15:jpcrd}
V.~A.~Yerokhin and V.~M.~Shabaev,
J.~Phys.~Chem.~Ref.~Data~\textbf{44}, 033103~(2015).
% 
\end{thebibliography}
%
%%%%%%%%%%%%%%%%%%%%%%%%%%%%%%%%%%%%%%%%%%%%%%%%%%%%%%%%%%%%%%%%%%%%%%%%
%

% 
%
%%%%%%%%%%%%%%%%%%%%%%%%%%%%%%%%%%%%%%%%%%%%%%%%%%%%%%%%%%%%%%%%%%%%%%%%
% 
%  TABLES
% 
%%%%%%%%%%%%%%%%%%%%%%%%%%%%%%%%%%%%%%%%%%%%%%%%%%%%%%%%%%%%%%%%%%%%%%%%
%
\newpage
%
% \begin{table}
\begin{longtable}{cr@{.}lr@{.}lr@{.}lr@{.}lr@{.}l}
% \centering
% \setlength{\tabcolsep}{1.2em}
\caption{The second-order interelectronic-interaction correction $\dgrecmzz$ to the magnetic nuclear recoil contribution to the ground-state $g$ factor of B-like ions evaluated within the Breit approximation. The results are expressed in terms of the function $C(\aZ)$ defined by Eq.~(\ref{eq:C_def}). 
\label{tab:C_magn}}
%\begin{ruledtabular}
% \begin{tabular}{cr@{.}lr@{.}lr@{.}lr@{.}lr@{.}l}
\vspace{1cm}\\
\hline 
\hline
$Z$
& \multicolumn{2}{c}{Coul}
& \multicolumn{2}{c}{CH}
& \multicolumn{2}{c}{PZ}
& \multicolumn{2}{c}{KS}
& \multicolumn{2}{c}{LDF}
% & \multicolumn{2}{c}{$C_\mathrm{Coul}$}
% & \multicolumn{2}{c}{$C_\mathrm{CH}$}
% & \multicolumn{2}{c}{$C_\mathrm{PZ}$} 
% & \multicolumn{2}{c}{$C_\mathrm{KS}$} 
% & \multicolumn{2}{c}{$C_\mathrm{LDF}$}
\\
\hline
   10  &  51&0 (3)    & $-$2&949 (10)  & $-$2&605 (10)  & $-$1&845 (10)  & $-$2&675 (10)  \\
   12  &  35&7 (2)    & $-$2&891 (9)   & $-$2&531 (8)   & $-$1&981 (9)   & $-$2&612 (9)   \\
   14  &  26&5 (1)    & $-$2&832 (8)   & $-$2&474 (8)   & $-$2&036 (8)   & $-$2&561 (8)   \\
   16  &  20&5 (1)    & $-$2&768 (8)   & $-$2&417 (8)   & $-$2&052 (8)   & $-$2&508 (8)   \\
   18  &  16&45 (8)   & $-$2&694 (8)   & $-$2&353 (8)   & $-$2&041 (8)   & $-$2&445 (8)   \\
   20  &  13&54 (6)   & $-$2&607 (8)   & $-$2&276 (8)   & $-$2&007 (8)   & $-$2&370 (8)   \\
   30  &   6&75 (3)   & $-$1&997 (7)   & $-$1&728 (7)   & $-$1&613 (7)   & $-$1&817 (7)   \\
   40  &   4&51 (1)   & $-$1&447 (7)   & $-$1&221 (7)   & $-$1&169 (7)   & $-$1&300 (7)   \\
   50  &   3&62 (1)   & $-$1&198 (7)   & $-$0&995 (7)   & $-$0&946 (7)   & $-$1&068 (7)   \\
   60  &   3&28 (1)   & $-$1&135 (6)   & $-$0&954 (6)   & $-$0&903 (6)   & $-$1&020 (6)   \\
   70  &   3&227 (8)  & $-$1&120 (6)   & $-$0&970 (6)   & $-$0&928 (6)   & $-$1&023 (6)   \\
   80  &   3&357 (7)  & $-$1&097 (6)   & $-$0&985 (6)   & $-$0&961 (6)   & $-$1&017 (6)   \\
   82  &   3&402 (7)  & $-$1&089 (6)   & $-$0&985 (6)   & $-$0&967 (6)   & $-$1&014 (6)   \\
   90  &   3&649 (6)  & $-$1&047 (6)   & $-$0&978 (5)   & $-$0&984 (6)   & $-$0&987 (6)   \\
   92  &   3&730 (5)  & $-$1&034 (5)   & $-$0&974 (5)   & $-$0&987 (5)   & $-$0&978 (5)   \\
\hline 
\hline            
% \end{tabular}
%\end{ruledtabular}
\end{longtable}
% \end{table}
%
% \vspace{5cm}
% \newpage
%
% \begin{table}
\begin{longtable}{ccrrrrrr}
% \centering
% \setlength{\tabcolsep}{0.5em}
\caption{The magnetic part of the nuclear recoil effect on the $g$ factor of B-like ions for the Coulomb and different effective screening potentials. The contributions of the zeroth ($A$), first ($B/Z$), and second ($C/Z^2$) orders in the interelectronic interaction are presented together with their sum, see Eqs.~(\ref{eq:C_def}) and (\ref{eq:dgrecm_k}). The ``spread'' is found as the maximal absolute difference between all screening-potential values. 
% The uncertainties (``unc'') are found as described in the text.
\label{tab:magn_total}}
%\begin{ruledtabular}
% \begin{tabular}{ccr@{.}lr@{.}lr@{.}lr@{.}lr@{.}lr@{.}lr@{.}l}
% \begin{tabular}{ccrrrrrrr}
% \begin{tabular}{ccrrrrrr}
\vspace{1cm}\\
\hline 
\hline
$Z$ & Term
& \multicolumn{1}{c}{Coul}
& \multicolumn{1}{c}{CH}
& \multicolumn{1}{c}{PZ}
& \multicolumn{1}{c}{KS}
& \multicolumn{1}{c}{LDF}
& \multicolumn{1}{c}{spread}
% & \multicolumn{1}{c}{unc}    
\\
\hline
  10  
& $A$           
& $-$0.77882 & $-$0.67297 & $-$0.67129 & $-$0.65265 & $-$0.66913  &  0.02032  \\
& $B/Z$         
&    0.18590 &    0.12193 &    0.11844 &    0.09351 &    0.11657  &           \\
& $A+B/Z$       
& $-$0.59292 & $-$0.55103 & $-$0.55285 & $-$0.55914 & $-$0.55257  &  0.00811  \\
& $C/Z^2$       
&    0.51001 & $-$0.02949 & $-$0.02605 & $-$0.01845 & $-$0.02675  &           \\
& $A+B/Z+C/Z^2$ 
& $-$0.08291 & $-$0.58052 & $-$0.57890 & $-$0.57759 & $-$0.57931  &  0.00293  \\
\cline{2-8}
& CI-DFS
& \multicolumn{5}{r}{$-$0.5750\phz} &  0.0043\phz  \\
\hline
  12  
& $A$           
& $-$0.77900 & $-$0.69748 & $-$0.69561 & $-$0.68230 & $-$0.69427  &  0.01518  \\
& $B/Z$         
&    0.15488 &    0.10182 &    0.09844 &    0.08183 &    0.09741  &           \\
& $A+B/Z$       
& $-$0.62412 & $-$0.59566 & $-$0.59716 & $-$0.60047 & $-$0.59686  &  0.00481  \\
& $C/Z^2$       
&    0.24793 & $-$0.02008 & $-$0.01758 & $-$0.01376 & $-$0.01814  &           \\
& $A+B/Z+C/Z^2$ 
& $-$0.37619 & $-$0.61573 & $-$0.61474 & $-$0.61422 & $-$0.61501  &  0.00151  \\
\cline{2-8}
& CI-DFS
& \multicolumn{5}{r}{$-$0.6116\phz} &  0.0035\phz  \\
\hline
  14
& $A$           
& $-$0.77922 & $-$0.71297 & $-$0.71119 & $-$0.70096 & $-$0.71031  &  0.01201  \\
& $B/Z$         
&    0.13273 &    0.08703 &    0.08406 &    0.07185 &    0.08344  &           \\
& $A+B/Z$       
& $-$0.64649 & $-$0.62594 & $-$0.62713 & $-$0.62911 & $-$0.62686  &  0.00317  \\
& $C/Z^2$       
&    0.13518 & $-$0.01445 & $-$0.01262 & $-$0.01039 & $-$0.01307  &           \\
& $A+B/Z+C/Z^2$ 
& $-$0.51132 & $-$0.64038 & $-$0.63975 & $-$0.63950 & $-$0.63993  &  0.00088  \\
\cline{2-8}
& CI-DFS
& \multicolumn{5}{r}{$-$0.6373\phz} &  0.0026\phz  \\
\hline
\pagebreak
\hline
  16
& $A$           
& $-$0.77947 & $-$0.72366 & $-$0.72204 & $-$0.71379 & $-$0.72143  &  0.00987  \\
& $B/Z$         
&    0.11610 &    0.07584 &    0.07327 &    0.06371 &    0.07289  &           \\
& $A+B/Z$       
& $-$0.66336 & $-$0.64783 & $-$0.64877 & $-$0.65007 & $-$0.64854  &  0.00224  \\
& $C/Z^2$       
&    0.08020 & $-$0.01081 & $-$0.00944 & $-$0.00802 & $-$0.00980  &           \\
& $A+B/Z+C/Z^2$ 
& $-$0.58316 & $-$0.65864 & $-$0.65821 & $-$0.65809 & $-$0.65834  &  0.00055  \\
\cline{2-8}
& CI-DFS
& \multicolumn{5}{r}{$-$0.6564\phz} &  0.0019\phz  \\
\hline
  18  
& $A$           
& $-$0.77975  & $-$0.73153  & $-$0.73007  & $-$0.72318  & $-$0.72964  &  0.00835  \\
& $B/Z$         
&    0.10318  &    0.06711  &    0.06489  &    0.05709  &    0.06466  &           \\
& $A+B/Z$       
& $-$0.67657  & $-$0.66442  & $-$0.66518  & $-$0.66609  & $-$0.66498  &  0.00167  \\
& $C/Z^2$       
&    0.05078  & $-$0.00832  & $-$0.00726  & $-$0.00630  & $-$0.00755  &           \\
& $A+B/Z+C/Z^2$ 
& $-$0.62579  & $-$0.67274  & $-$0.67244  & $-$0.67239  & $-$0.67253  &  0.00035  \\
\cline{2-8}
& CI-DFS
& \multicolumn{5}{r}{$-$0.6711\phz} &  0.0014\phz  \\
\hline
  20  
& $A$           
& $-$0.78006 & $-$0.73759 & $-$0.73628 & $-$0.73038 & $-$0.73598  &  0.00721  \\
& $B/Z$         
&    0.09283 &    0.06013 &    0.05820 &    0.05163 &    0.05806  &           \\
& $A+B/Z$       
& $-$0.68724 & $-$0.67746 & $-$0.67808 & $-$0.67875 & $-$0.67792  &  0.00129  \\
& $C/Z^2$       
&    0.03386 & $-$0.00652 & $-$0.00569 & $-$0.00502 & $-$0.00592  &           \\
& $A+B/Z+C/Z^2$ 
& $-$0.65338 & $-$0.68398 & $-$0.68377 & $-$0.68377 & $-$0.68384  &  0.00021  \\
\cline{2-8}
& CI-DFS
& \multicolumn{5}{r}{$-$0.6828\phz} &  0.0011\phz  \\
\hline
  30  
& $A$           
& $-$0.78215 & $-$0.75533 & $-$0.75456 & $-$0.75116 & $-$0.75452  &  0.00417  \\
& $B/Z$         
&    0.06178 &    0.03924 &    0.03820 &    0.03459 &    0.03825  &           \\
& $A+B/Z$       
& $-$0.72037 & $-$0.71609 & $-$0.71636 & $-$0.71657 & $-$0.71627  &  0.00048  \\
& $C/Z^2$       
&    0.00750 & $-$0.00222 & $-$0.00192 & $-$0.00179 & $-$0.00202  &           \\
& $A+B/Z+C/Z^2$ 
& $-$0.71288 & $-$0.71830 & $-$0.71828 & $-$0.71836 & $-$0.71829  &  0.00008  \\
\cline{2-8}
& CI-DFS
& \multicolumn{5}{r}{$-$0.7181\phz} &  0.0002\phz  \\
\hline
\pagebreak
\hline
  40  
& $A$           
& $-$0.78509 & $-$0.76521 & $-$0.76478 & $-$0.76240 & $-$0.76479  &  0.00281  \\
& $B/Z$         
&    0.04625 &    0.02880 &    0.02823 &    0.02576 &    0.02829  &           \\
& $A+B/Z$       
& $-$0.73884 & $-$0.73641 & $-$0.73655 & $-$0.73664 & $-$0.73650  &  0.00023  \\
& $C/Z^2$       
&    0.00282 & $-$0.00090 & $-$0.00076 & $-$0.00073 & $-$0.00081  &           \\
& $A+B/Z+C/Z^2$ 
& $-$0.73602 & $-$0.73731 & $-$0.73731 & $-$0.73737 & $-$0.73731  &  0.00006  \\
\cline{2-8}
& CI-DFS
& \multicolumn{5}{r}{$-$0.73734} &  0.00003  \\
\hline
  50  
& $A$           
& $-$0.788889 & $-$0.772855 & $-$0.772677 & $-$0.770847 & $-$0.772658  &  0.002008  \\
& $B/Z$         
&    0.036958 &    0.022512 &    0.022248 &    0.020370 &    0.022262  &           \\
& $A+B/Z$       
& $-$0.751931 & $-$0.750342 & $-$0.750429 & $-$0.750477 & $-$0.750397  &  0.000135  \\
& $C/Z^2$       
&    0.001447 & $-$0.000479 & $-$0.000398 & $-$0.000378 & $-$0.000427  &           \\
& $A+B/Z+C/Z^2$ 
& $-$0.750484 & $-$0.750821 & $-$0.750827 & $-$0.750855 & $-$0.750824  &  0.000034  \\
\cline{2-8}
& CI-DFS
& \multicolumn{5}{r}{$-$0.75085\phz} &  0.00003\phz  \\
\hline
  60  
& $A$           
& $-$0.793550 & $-$0.779912 & $-$0.779935 & $-$0.778452 & $-$0.779858  &  0.001483  \\
& $B/Z$         
&    0.030798 &    0.018296 &    0.018263 &    0.016754 &    0.018208  &           \\
& $A+B/Z$       
& $-$0.762752 & $-$0.761616 & $-$0.761672 & $-$0.761698 & $-$0.761650  &  0.000082  \\
& $C/Z^2$       
&    0.000911 & $-$0.000315 & $-$0.000265 & $-$0.000251 & $-$0.000283  &           \\
& $A+B/Z+C/Z^2$ 
& $-$0.761841 & $-$0.761931 & $-$0.761937 & $-$0.761949 & $-$0.761933  &  0.000018  \\
\cline{2-8}
& CI-DFS
& \multicolumn{5}{r}{$-$0.76195\phz} &  0.00001\phz  \\
\hline
  70  
& $A$           
& $-$0.799063 & $-$0.787029 & $-$0.787229 & $-$0.785988 & $-$0.787074  &  0.001241  \\
& $B/Z$         
&    0.026438 &    0.015272 &    0.015432 &    0.014177 &    0.015293  &           \\
& $A+B/Z$       
& $-$0.772625 & $-$0.771757 & $-$0.771797 & $-$0.771811 & $-$0.771781  &  0.000054  \\
& $C/Z^2$       
&    0.000659 & $-$0.000229 & $-$0.000198 & $-$0.000189 & $-$0.000209  &           \\
& $A+B/Z+C/Z^2$ 
& $-$0.771966 & $-$0.771986 & $-$0.771995 & $-$0.772000 & $-$0.771990  &  0.000014  \\
\cline{2-8}
& CI-DFS
& \multicolumn{5}{r}{$-$0.77200\phz} &  0.00001\phz  \\
\hline
\pagebreak
\hline
  80  
& $A$           
& $-$0.805355 & $-$0.794453 & $-$0.794819 & $-$0.793762 & $-$0.794572  &  0.001057  \\
& $B/Z$         
&    0.023200 &    0.012998 &    0.013333 &    0.012268 &    0.013098  &           \\
& $A+B/Z$       
& $-$0.782155 & $-$0.781454 & $-$0.781487 & $-$0.781494 & $-$0.781474  &  0.000040  \\
& $C/Z^2$       
&    0.000525 & $-$0.000171 & $-$0.000154 & $-$0.000150 & $-$0.000159  &           \\
& $A+B/Z+C/Z^2$ 
& $-$0.781630 & $-$0.781625 & $-$0.781641 & $-$0.781644 & $-$0.781633  &  0.000019  \\
\cline{2-8}
& CI-DFS
& \multicolumn{5}{r}{$-$0.78163\phz} &  0.00000\phz  \\
\hline
  82  
& $A$           
& $-$0.806700 & $-$0.795984 & $-$0.796383 & $-$0.795358 & $-$0.796116  &  0.001025  \\
& $B/Z$         
&    0.022650 &    0.012610 &    0.012977 &    0.011946 &    0.012723  &           \\
& $A+B/Z$       
& $-$0.784050 & $-$0.783374 & $-$0.783406 & $-$0.783412 & $-$0.783393  &  0.000038  \\
& $C/Z^2$       
&    0.000506 & $-$0.000162 & $-$0.000147 & $-$0.000144 & $-$0.000151  &           \\
& $A+B/Z+C/Z^2$ 
& $-$0.783544 & $-$0.783536 & $-$0.783553 & $-$0.783556 & $-$0.783544  &  0.000020  \\
\cline{2-8}
& CI-DFS
& \multicolumn{5}{r}{$-$0.78354\phz} &  0.00000\phz  \\
\hline
  90  
& $A$           
& $-$0.812340 & $-$0.802283 & $-$0.802811 & $-$0.801906 & $-$0.802461  &  0.000905  \\
& $B/Z$         
&    0.020686 &    0.011227 &    0.011723 &    0.010815 &    0.011386  &           \\
& $A+B/Z$       
& $-$0.791654 & $-$0.791056 & $-$0.791088 & $-$0.791091 & $-$0.791075  &  0.000035  \\
& $C/Z^2$       
&    0.000451 & $-$0.000129 & $-$0.000121 & $-$0.000122 & $-$0.000122  &           \\
& $A+B/Z+C/Z^2$ 
& $-$0.791203 & $-$0.791185 & $-$0.791209 & $-$0.791213 & $-$0.791197  &  0.000028  \\
\cline{2-8}
& CI-DFS
& \multicolumn{5}{r}{$-$0.79120\phz} &  0.00000\phz  \\
\hline
  92  
& $A$           
& $-$0.813803 & $-$0.803894 & $-$0.804455 & $-$0.803577 & $-$0.804082  &  0.000878  \\
& $B/Z$         
&    0.020244 &    0.010918 &    0.011446 &    0.010566 &    0.011087  &           \\
& $A+B/Z$       
& $-$0.793559 & $-$0.792976 & $-$0.793009 & $-$0.793011 & $-$0.792995  &  0.000035  \\
& $C/Z^2$       
&    0.000441 & $-$0.000122 & $-$0.000115 & $-$0.000117 & $-$0.000115  &           \\
& $A+B/Z+C/Z^2$ 
& $-$0.793118 & $-$0.793098 & $-$0.793124 & $-$0.793128 & $-$0.793110  &  0.000030  \\
\cline{2-8}
& CI-DFS
& \multicolumn{5}{r}{$-$0.79312\phz} &  0.00001\phz  \\
\hline
%   20  & $A$           &  $-$0.77596  &  $-$0.73398  &  $-$0.73259 &  $-$0.72669  &  $-$0.73233 \\
%       & $B/Z$         &     0.09235  &     0.06012  &     0.05810 &     0.05152  &     0.05801 \\
%       & $A+B/Z$       &  $-$0.68361  &  $-$0.67386  &  $-$0.67449 &  $-$0.67517  &  $-$0.67432 \\
%       & $C/Z^2$       &     0.09235  &     0.06012  &     0.05810 &     0.05152  &     0.05801 \\
%       & $A+B/Z+C/Z^2$ &  $-$0.68361  &  $-$0.67386  &  $-$0.67449 &  $-$0.67517  &  $-$0.67432 \\
% \hline
%   92  & $A$           &  $-$0.81380  &  $-$0.80389  &  $-$0.80446  &  $-$0.80358  &  $-$0.80408  &  0.00088  &  \\
%       & $B/Z$         &     0.02024  &     0.01092  &     0.01145  &     0.01057  &     0.01109  &  &  \\
%       & $A+B/Z$       &  $-$0.79356  &  $-$0.79297  &  $-$0.79301  &  $-$0.79301  &  $-$0.79299  &  0.00003  &  0.00033  \\
%       & $C/Z^2$       &     0.00044  &  $-$0.00012  &  $-$0.00012  &  $-$0.00012  &  $-$0.00012  &  &  \\
%       & $A+B/Z+C/Z^2$ &  $-$0.79312  &  $-$0.79310  &  $-$0.79312  &  $-$0.79312  &  $-$0.79311  &  0.00003  &  \\
\hline 
\hline            
% \end{tabular}
%\end{ruledtabular}
\end{longtable}
% \end{table}
%
\newpage
%
% \begin{table}
\begin{longtable}{ccrrrrr}
% \centering
% \setlength{\tabcolsep}{0.5em}
\caption{The nonmagnetic part of the nuclear recoil effect on the $g$ factor of B-like ions for the Coulomb and different effective screening potentials. The contributions of the zeroth ($A$) and first ($B/Z$) orders in the interelectronic interaction are presented together with their sum. 
\label{tab:nonmagn_total}}
%\begin{ruledtabular}
% \begin{tabular}{ccr@{.}lr@{.}lr@{.}lr@{.}lr@{.}lr@{.}lr@{.}l}
% \begin{tabular}{ccrrrrr}
\vspace{1cm}\\
\hline 
\hline
$Z$& Term &
     \multicolumn{1}{c}{Coul}  & 
     \multicolumn{1}{c}{CH}    &
     \multicolumn{1}{c}{PZ}    &
     \multicolumn{1}{c}{KS}    &
     \multicolumn{1}{c}{LDF}   \\
\hline
  10  & $A$           &     0.001010  &     0.000784  &     0.000817  &     0.000819  &     0.000797  \\
      & $B/Z$         &  $-$0.000226  &  $-$0.000035  &  $-$0.000073  &  $-$0.000092  &  $-$0.000047  \\
      & $A+B/Z$       &     0.000784  &     0.000749  &     0.000744  &     0.000727  &     0.000750  \\
\hline
  12  & $A$           &     0.001458  &     0.001180  &     0.001221  &     0.001221  &     0.001197  \\
      & $B/Z$         &  $-$0.000274  &  $-$0.000029  &  $-$0.000076  &  $-$0.000092  &  $-$0.000047  \\
      & $A+B/Z$       &     0.001185  &     0.001151  &     0.001145  &     0.001130  &     0.001150  \\
\hline
  14  & $A$           &     0.001990  &     0.001660  &     0.001710  &     0.001708  &     0.001681  \\
      & $B/Z$         &  $-$0.000322  &  $-$0.000025  &  $-$0.000080  &  $-$0.000093  &  $-$0.000047  \\
      & $A+B/Z$       &     0.001668  &     0.001635  &     0.001629  &     0.001615  &     0.001634  \\
\hline
  16  & $A$           &     0.002608  &     0.002225  &     0.002283  &     0.002279  &     0.002250  \\
      & $B/Z$         &  $-$0.000372  &  $-$0.000021  &  $-$0.000086  &  $-$0.000096  &  $-$0.000048  \\
      & $A+B/Z$       &     0.002236  &     0.002204  &     0.002197  &     0.002184  &     0.002202  \\
\hline
  18  & $A$           &     0.003313  &     0.002877  &     0.002943  &     0.002938  &     0.002906  \\
      & $B/Z$         &  $-$0.000424  &  $-$0.000019  &  $-$0.000092  &  $-$0.000100  &  $-$0.000049  \\
      & $A+B/Z$       &     0.002889  &     0.002858  &     0.002851  &     0.002838  &     0.002856  \\
\hline
  20  & $A$           &     0.004107  &     0.003616  &     0.003692  &     0.003685  &     0.003649  \\
      & $B/Z$         &  $-$0.000477  &  $-$0.000017  &  $-$0.000099  &  $-$0.000105  &  $-$0.000052  \\
      & $A+B/Z$       &     0.003630  &     0.003600  &     0.003592  &     0.003580  &     0.003598  \\
\hline
  30  & $A$           &     0.009500  &     0.008718  &     0.008843  &     0.008830  &     0.008774  \\
      & $B/Z$         &  $-$0.000778  &  $-$0.000022  &  $-$0.000156  &  $-$0.000154  &  $-$0.000081  \\
      & $A+B/Z$       &     0.008722  &     0.008696  &     0.008687  &     0.008676  &     0.008693  \\
\hline
\pagebreak
\hline
  40  & $A$           &     0.017568  &     0.016450  &     0.016640  &     0.016622  &     0.016534  \\
      & $B/Z$         &  $-$0.001156  &  $-$0.000057  &  $-$0.000258  &  $-$0.000251  &  $-$0.000144  \\
      & $A+B/Z$       &     0.016413  &     0.016393  &     0.016382  &     0.016371  &     0.016390  \\
\hline
  50  & $A$           &     0.028922  &     0.027401  &     0.027679  &     0.027659  &     0.027520  \\
      & $B/Z$         &  $-$0.001637  &  $-$0.000130  &  $-$0.000421  &  $-$0.000411  &  $-$0.000253  \\
      & $A+B/Z$       &     0.027285  &     0.027271  &     0.027258  &     0.027248  &     0.027267  \\
\hline
  60  & $A$           &     0.044507  &     0.042488  &     0.042889  &     0.042872  &     0.042653  \\
      & $B/Z$         &  $-$0.002255  &  $-$0.000244  &  $-$0.000661  &  $-$0.000655  &  $-$0.000414  \\
      & $A+B/Z$       &     0.042252  &     0.042244  &     0.042228  &     0.042217  &     0.042239  \\
\hline
  70  & $A$           &     0.065792  &     0.063140  &     0.063718  &     0.063714  &     0.063369  \\
      & $B/Z$         &  $-$0.003043  &  $-$0.000396  &  $-$0.000993  &  $-$0.001000  &  $-$0.000630  \\
      & $A+B/Z$       &     0.062749  &     0.062745  &     0.062724  &     0.062714  &     0.062739  \\
\hline
  80  & $A$           &     0.095116  &     0.091640  &     0.092479  &     0.092506  &     0.091958  \\
      & $B/Z$         &  $-$0.004038  &  $-$0.000565  &  $-$0.001431  &  $-$0.001470  &  $-$0.000891  \\
      & $A+B/Z$       &     0.091078  &     0.091075  &     0.091048  &     0.091036  &     0.091067  \\
\hline
  82  & $A$           &     0.102243  &     0.098573  &     0.099480  &     0.099516  &     0.098914  \\
      & $B/Z$         &  $-$0.004262  &  $-$0.000596  &  $-$0.001531  &  $-$0.001580  &  $-$0.000945  \\
      & $A+B/Z$       &     0.097981  &     0.097978  &     0.097949  &     0.097936  &     0.097969  \\
\hline
  90  & $A$           &     0.136238  &     0.131675  &     0.132915  &     0.133006  &     0.132124  \\
      & $B/Z$         &  $-$0.005227  &  $-$0.000672  &  $-$0.001950  &  $-$0.002055  &  $-$0.001133  \\
      & $A+B/Z$       &     0.131011  &     0.131003  &     0.130964  &     0.130950  &     0.130991  \\
\hline
  92  & $A$           &     0.146358  &     0.141539  &     0.142882  &     0.142993  &     0.142022  \\
      & $B/Z$         &  $-$0.005477  &  $-$0.000670  &  $-$0.002054  &  $-$0.002180  &  $-$0.001165  \\
      & $A+B/Z$       &     0.140880  &     0.140869  &     0.140828  &     0.140813  &     0.140856  \\
\hline 
\hline            
% \end{tabular}
%\end{ruledtabular}
\end{longtable}
% \end{table}
%
\newpage
%
% \begin{table}
\begin{longtable}{cr@{.}lr@{.}lr@{.}l}
% \centering
% \setlength{\tabcolsep}{1.2em}
\caption{The Breit, QED, and total nuclear recoil contributions of the first order in the electron-to-nucleus mass ratio $m/M$ to the ground-state $g$ factor of B-like ions expressed in terms of the function $F(\alpha Z)$ defined by Eq.~(\ref{eq:F_def}). 
% The uncertainties are determined according to the procedure described in the text. 
\label{tab:total}}
%\begin{ruledtabular}
% \begin{tabular}{cr@{.}lr@{.}lr@{.}l}
\vspace{1cm}\\
\hline 
\hline
$Z$
& \multicolumn{2}{c}{$\FBreit(\aZ)$}
& \multicolumn{2}{c}{$\FQED(\aZ)$}
& \multicolumn{2}{c}{$\Frec(\aZ)$}
\\
\hline
   10  &   $-$0&5743 (43)   &   0&000013 (4) &    $-$0&5743 (43) \\
   12  &   $-$0&6104 (35)   &   0&000024 (5) &    $-$0&6104 (35) \\
   14  &   $-$0&6357 (26)   &   0&000039 (7) &    $-$0&6357 (26) \\
   16  &   $-$0&6542 (19)   &   0&000060 (10)&    $-$0&6541 (19) \\
   18  &   $-$0&6682 (14)   &   0&000087 (12)&    $-$0&6681 (14) \\
   20  &   $-$0&6792 (11)   &   0&00012 (1)  &    $-$0&6790 (11) \\
   30  &   $-$0&7095 (2)    &   0&00041 (3)  &    $-$0&7090 (2)  \\
   40  &   $-$0&7210 (1)    &   0&00094 (5)  &    $-$0&7200 (1)  \\
   50  &   $-$0&7236 (1)    &   0&0018 (1)   &    $-$0&7218 (1)  \\
   60  &   $-$0&7197 (1)    &   0&0029 (1)   &    $-$0&7168 (1)  \\
   70  &   $-$0&7093 (2)    &   0&0045 (1)   &    $-$0&7048 (2)  \\
   80  &   $-$0&6906 (3)    &   0&0070 (2)   &    $-$0&6836 (4)  \\
   82  &   $-$0&6856 (4)    &   0&0077 (2)   &    $-$0&6779 (5)  \\
   90  &   $-$0&6602 (10)   &   0&0121 (3)   &    $-$0&6481 (11) \\
   92  &   $-$0&6523 (13)   &   0&0138 (4)   &    $-$0&6384 (13) \\  
\hline 
\hline            
% \end{tabular}
%\end{ruledtabular}
\end{longtable}
% \end{table}
%
\newpage
% 
% \begin{table}
\begin{longtable}{cr@{.}lr@{.}l}
% \centering
% \setlength{\tabcolsep}{1.2em}
\caption{Nuclear recoil contribution of the first order in the electron-to-nucleus mass ratio $m/M$ to the ground-state $g$ factor of selected B-like ions in the range $Z=10-60$.
\label{tab:total_ions}}
%\begin{ruledtabular}
% \begin{tabular}{cr@{.}lr@{.}l}
\vspace{1cm}\\
\hline 
\hline
Ion
& \multicolumn{2}{c}{$(m/M) \cdot 10^6$}        
% & \multicolumn{2}{c}{$\dgrecBreit \cdot 10^6$}          
% & \multicolumn{2}{c}{$\dgrecQED \cdot 10^6$}
& \multicolumn{2}{c}{$\dgrec \cdot 10^6$}
\\
\hline
${}^{20}_{10}$Ne$^{5+}$
         &     27&4469
         &  $-$15&76 (12)
\\
${}^{24}_{12}$Mg$^{7+}$
         &     22&8780
         &  $-$13&96 (8)
\\
${}^{28}_{14}$Si$^{9+}$
         &     19&6137
         &  $-$12&47 (5)
\\
${}^{32}_{16}$S$^{11+}$
         &     17&1628
         &  $-$11&226 (33)
\\
${}^{40}_{18}$Ar$^{13+}$
         &     13&7308
         &   $-$9&174 (19)
\\
${}^{40}_{20} \text{Ca}^{15+}$  
         &     13&7311
         &   $-$9&323 (15)
\\
${}^{48}_{20} \text{Ca}^{15+}$  
         &     11&4427
         &   $-$7&770 (13)
\\
${}^{120}_{50} \text{Sn}^{45+}$  
         &      4&57628     
         &   $-$3&3033 (5)
\\
${}^{142}_{60} \text{Nd}^{55+}$  
         &      3&86665     
         &   $-$2&7717 (6)
\\
\hline 
\hline            
% \end{tabular}
%\end{ruledtabular}
\end{longtable}
% \end{table}
%
% 
\end{document}